\documentclass[aps,prx,twocolumn,preprintnumbers,,superscriptaddress,nofootinbib,reprint]{revtex4-2}
\pdfoutput=1
\usepackage{amsmath,amssymb,braket,bm}
\usepackage{graphicx}
\usepackage{hyperref}
\usepackage{physics}
\usepackage{color,xcolor}

\usepackage[normalem]{ulem}

\begin{document}
\preprint{IQuS@UW-21-127, NT@UW-26-11}
\title{Quantum Simulation of Nucleon-Antinucleon Interaction in Large-$N$ QCD$_2$ \\ on an IBM Quantum Nighthawk Processor}

\author{Cameron V. Cogburn}
\email{cogbuc@rpi.edu}
\affiliation{Future of Computing Institute, Rensselaer Polytechnic Institute, Troy, NY 12180, USA}

\author{Sebastian Grieninger}
\email{segrie@uw.edu}
\affiliation{InQubator for Quantum Simulation (IQuS), Department of Physics, University of Washington, Seattle, WA 98195, USA}

\author{Dmitri E. Kharzeev}
\email{dmitri.kharzeev@stonybrook.edu}
\affiliation{
 Center for Nuclear Theory, Department of Physics and Astronomy,
Stony Brook University, Stony Brook, New York 11794-3800, USA
}
\affiliation{Energy and Photon Sciences Directorate, Condensed Matter and Materials Sciences Division,
Brookhaven National Laboratory, Upton, New York 11973-5000, USA}

\date{\today}

\begin{abstract}
We report a quantum simulation of the nucleon--antinucleon interaction in large-$N$ two-dimensional quantum chromodynamics (QCD$_2$) on the IBM Quantum Nighthawk processor. In the large-$N$ limit, QCD$_2$ admits a bosonized description in which baryons emerge as topological solitons (kinks) of an effective mesonic field theory, providing a controlled, nonperturbative framework for baryon--antibaryon dynamics.

We formulate the problem by mapping the continuum bosonized Hamiltonian to a spin-chain representation equivalent to an XXZ model with anisotropy set by the QCD parameters. In this mapping, nucleon and antinucleon states correspond to kink and antikink excitations, respectively, while their interaction is encoded in the spin correlations of the chain. Using Jordan--Wigner encoding, we implement the resulting XXZ Hamiltonian on a finite set of qubits and realize it via a variational ground state ansatz and postselected nonunitary disorder operator insertions optimized for the Nighthawk architecture.
We then show the kink--antikink interaction potential built from the conditional energies of these nonunitary string operators can be robustly extracted from the quantum hardware due to structured error cancelation.
The resulting potential exhibits the expected attractive behavior.
The quantum simulation results are benchmarked against exact diagonalization, ideal statevector evaluation showing good agreement. To connect the device result to the continuum field theory, we extract the potential in the continuum limit using large-$L$ matrix product state calculations.

\end{abstract}

\maketitle
\section{Introduction}

Understanding the interaction between a nucleon and an antinucleon occupies a central place in the nonperturbative dynamics of the strong interaction. The nucleon--antinucleon interaction probes some of the key properties of quantum chromodynamics (QCD): confinement, chiral symmetry breaking, and topology. In particular, the interplay between attraction, repulsion, and annihilation channels in the baryon--antibaryon sector (see, e.g., \cite{Dover1992,Kang:2013uia}) reflects the collective dynamics of many degrees of freedom and remains difficult to address quantitatively using classical computational methods~\cite{Luscher:1986pf,Luscher:1990ux,Beane:2010em}.

Theoretical progress can be achieved by considering controlled limits of QCD that retain its essential nonperturbative features while allowing for analytical insight and systematic approximations. One such limit is QCD in two spacetime dimensions at a large number of colors, $N$, commonly referred to as large-$N$ QCD$_2$ \cite{tHooft:1974pnl,Callan:1975ps,Witten:1983ar,Frishman:1992mr,Gross:1995bp}. Despite its reduced dimensionality, QCD$_2$ shares key characteristics with four-dimensional QCD, including confinement and a nontrivial vacuum structure. Remarkably, in the large-$N$ limit the theory admits an exact bosonized description, in which baryons and antibaryons emerge as topological solitons of an effective scalar field theory~\cite{Frishman:1992mr}. This mapping renders the nucleon--antinucleon interaction amenable to a transparent and physically intuitive formulation.

A cornerstone of this correspondence is the equivalence between large-$N$ QCD$_2$ with massive fermions and the sine-Gordon (SG) model. In this framework, baryons and antibaryons are identified with soliton and antisoliton excitations, while their interactions are encoded in the nonlinear dynamics of the SG field. The nucleon--antinucleon potential, bound states, and annihilation processes translate into soliton--antisoliton scattering and annihilation in the SG model. This correspondence not only clarifies the topological origin of baryon number but also establishes a direct bridge between non-Abelian gauge dynamics and integrable quantum field theories.

Crucially for the present work, the sine-Gordon model~\cite{Coleman:1974bu,Mandelstam:1975hb,Dashen:1975hd,Faddeev:1977rm,Zamolodchikov:1978xm,Rajaraman:1982is} itself admits a well-known mapping onto a lattice spin system. Through standard bosonization and Jordan--Wigner transformations, the SG field theory can be discretized and mapped onto the anisotropic spin-$\tfrac{1}{2}$ XXZ chain in a longitudinal field \cite{Affleck1999,Oshikawa1997,Luther:1975wr,Haldane:1981zza,Giamarchi:2003ooa,Korepin:1993kvr,Faddeev:1979gh}. In this representation, solitons correspond to domain wall excitations, while soliton--antisoliton interactions are encoded in the spin--spin couplings and anisotropy parameters of the XXZ Hamiltonian. The resulting model is local, finite-dimensional, and naturally formulated in terms of qubits, making it ideally suited for quantum simulation~\cite{Bauer:2022hpo,Bauer:2023qgm,Ciavarella:2024fzw,Funcke:2023jbq,DiMeglio:2023nsa}.

In this paper, we present a proof-of-principle quantum simulation of nucleon--antinucleon interactions in large-$N$ QCD$_2$ by exploiting a sequence of well-controlled theoretical mappings:
\begin{equation*}
\text{QCD}_2 \ (N\!\to\!\infty)
\;\longrightarrow\;
\text{sine-Gordon}
\;\longrightarrow\;
\text{XXZ spin chain}.
\end{equation*}

We implement the resulting XXZ Hamiltonian on an IBM Quantum Nighthawk processor and explore the potential acting between the soliton and antisoliton states, corresponding to nucleon--antinucleon potential in large-N QCD$_2$.

It is instructive to contrast our approach with the challenges that arise 
when attempting to extract nucleon--antinucleon (or nucleon--nucleon) potentials in lattice QCD 
at finite baryon mass~\cite{Nicholson:2021zwi,Amarasinghe:2021lqa,NPLQCD:2012mex,NPLQCD:2013bqy,Iritani:2018vfn,Ishii:2006ec,Iritani:2016jie,Beane:2017edf,Tews:2022yfb,Green:2021sxb,Green:2025rel}. The fundamental difficulty is that the operator of the second baryon, when applied to construct the two-baryon state, may not act on an energy eigenstate. 
As a result, the state may contain contributions from all eigenstates at finite momentum, and these contributions cannot be systematically removed at any finite baryon mass. This may cause scheme-dependence of the potential.

In the large-$N$ limit studied here, this problem is absent by 
construction. Through the bosonization mapping from large-$N$ QCD$_2$ 
to the sine-Gordon model, the sine-Gordon coupling takes the 
value~\cite{Faddeev:1977rm,Frishman:1992mr,Florio:2022uvd} $\beta^2 = 4\pi/N$. The classical baryon 
mass, identified with the sine-Gordon soliton mass, scales 
as $M_B \propto 1/\beta^2 \propto N$, and 
therefore diverges in the $N \to \infty$ limit resulting in infinitely heavy, static baryons. 
In this limit the nucleon--antinucleon potential reduces to a static 
potential between fixed sources~\cite{Witten:1979kh,Kaplan:1995yg,Kaplan:1996rk}. Since the sources 
carry no momentum, the eigenstate contamination problem is removed 
entirely and the potential is well-defined and scheme-independent. 
The large-$N$ limit therefore provides the unique theoretically clean 
setting in which the concept of an interbaryon potential is well-posed~\cite{Kaplan:1995yg,Kaplan:1996rk,Beane:2002ab},
and our quantum simulation operates entirely within this regime.

\section{Model and observables}
\label{sec:model}

\subsection{From large-$N$ QCD$_2$ to sine-Gordon}
\label{sec:qcd2_to_sg}

We consider $1\!+\!1$ dimensional QCD with gauge group $\mathrm{SU}(N)$ and $N_f=1$ massive fundamental fermion,
\begin{equation}
\mathcal{L}_{\mathrm{QCD}_2}
=
-\frac{1}{2}\,\mathrm{tr}\,F_{\mu\nu}F^{\mu\nu}
+\bar{\psi}\,( i\gamma^\mu D_\mu - m)\psi,
\end{equation}
where $D_\mu=\partial_\mu-i g A_\mu$, and take the 't~Hooft limit $N\to\infty$ at fixed $\lambda \equiv g^2 N$.
In two dimensions the gauge field has no propagating degrees of freedom; integrating it out yields an instantaneous confining interaction among color charges. In the large-$N$ limit, bosonization leads to an equivalent description in terms of a compact scalar field $\phi(x,t)$ whose dynamics is governed by a sine-Gordon (SG) model. In this correspondence, the conserved baryon number is identified with the topological charge of $\phi$,
\begin{equation}
B=\frac{\beta}{2\pi}
\int dx\,\partial_x \phi(x,t)=\frac{\beta}{2\pi}
\Big[\phi(+\infty,t)-\phi(-\infty,t)\Big],
\label{eq:baryon_top_charge}
\end{equation}
so that a baryon (antibaryon) corresponds to a soliton (antisoliton) configuration interpolating between adjacent vacua of the SG potential, each separated by $\Delta\phi = 2\pi/\beta$, yielding $B=\pm 1$.

We thus take as our effective continuum description
\begin{equation}
\mathcal{L}_{\mathrm{SG}}
=
\frac{1}{2}(\partial_\mu \phi)(\partial^\mu \phi)
+\frac{\alpha}{\beta^2}\cos\!\big(\beta \phi\big),
\qquad
\mu=0,1,
\label{eq:sg_lagrangian}
\end{equation}
with parameters $(\alpha,\beta)$ determined by $(m,\lambda)$ in the underlying gauge theory. The vacuum manifold consists of the degenerate minima $\phi = 2\pi k/\beta$, $k\in\mathbb{Z}$, and the soliton/antisoliton excitations carry $B=\pm1$ via \eqref{eq:baryon_top_charge}. The soliton mass $M_s$ and the soliton--antisoliton scattering data are then controlled by the SG coupling $\beta$ (or, equivalently, by the XXZ anisotropy in the lattice representation below).

\subsection{From sine-Gordon to the lattice XXZ chain}
\label{sec:sg_to_xxz}

To enable gate-based quantum simulation we employ a lattice regularization in which the SG field theory maps to a one-dimensional spin-$\tfrac{1}{2}$ XXZ chain (equivalently, an interacting fermion chain via Jordan--Wigner). Concretely, the SG model can be viewed as the continuum limit of a lattice model with Hamiltonian
\begin{equation}
H_{\mathrm{XXZ}}
=
J\sum_{n=1}^{L-1}\Big(
\sigma_n^x\sigma_{n+1}^x+\sigma_n^y\sigma_{n+1}^y
+\Delta\,\sigma_n^z\sigma_{n+1}^z
\Big)
+\sum_{n=1}^{L} h_n\,\sigma_n^z,
\label{eq:xxz_hamiltonian}
\end{equation}
where $J$ sets the overall energy scale, $\Delta$ is the anisotropy parameter, and $h_n$ is an effective longitudinal field (used to create localized soliton profiles corresponding to massive quark). The continuum SG coupling is related to the lattice anisotropy by
the Bethe-ansatz Luttinger parameter~\cite{Korepin:1993kvr,Giamarchi:2003ooa} (see Appendix~\ref{sec:continuum_mapping}),
\begin{equation}
\beta^2 = 4\pi K(\Delta), \qquad K(\Delta) = \frac{\pi}{2(\pi - \arccos\Delta)},
\label{eq:beta_delta_relation}
\end{equation}
which is exact in the critical regime $-1<\Delta\le 1$.
\footnote{In practice, we treat the mapping as defining a one-parameter family of SG theories realized by the XXZ chain; the physical matching to large-$N$ QCD$_2$ fixes the SG coupling and soliton mass scale.}

In the spin language, SG solitons correspond to domain wall (kink) excitations in $\sigma^z$ between regions polarized along $\pm z$. The topological charge $B$ maps to the net change of the boson field, which on the lattice is represented by the number and orientation of domain walls (or, equivalently, by the fermion number in the Jordan--Wigner representation). This identification allows us to prepare soliton and antisoliton states as localized kink/antikink wavepackets in the XXZ chain and to study their interaction potential.

\subsection{Mapping to sine-Gordon theory and soliton mass gap}
\label{sec:SG_mapping}

We consider the spin-$\tfrac12$ XXZ chain with a purely staggered longitudinal field,
\begin{equation}
H
=
J\sum_{n=1}^{L-1}\Big(
\sigma_n^x\sigma_{n+1}^x+\sigma_n^y\sigma_{n+1}^y
+\Delta\,\sigma_n^z\sigma_{n+1}^z
\Big)
+
h_s\sum_{n=1}^{L}(-1)^n \sigma_n^z ,
\label{eq:XXZ_staggered}
\end{equation}
in the critical regime $-1<\Delta\le 1$ and for small $|h_s|$.
At long wavelengths this model admits a controlled bosonized description, which we now relate to the large-$N$ one-flavor $\mathrm{QCD}_2$ effective theory.

Excitations of the spin chain are created by nonlocal ``disorder" operators~\cite{Mandelstam:1975hb,Fradkin:1978th} that generate domain walls on the N\'{e}el order. On the lattice, the kink and antikink operators can be written as string operators~\cite{Jordan:1928wi,Lieb:1961fr,Schultz:1964fv}:
\begin{equation}
\mu_j^+ =
\exp\!\left( i\pi \sum_{l<j} ( S_l^z - m ) \right) S_j^+,
\label{eq:mu}
\end{equation}
where $m=\langle S^z\rangle=0$ at half filling and $\mu_j^-$ follows analogous.
For spin-$\tfrac{1}{2}$ degrees of freedom, up to an overall phase, the exponential string reduces to a product of Pauli operators,
\(
\exp(i\pi S_l^z)\propto \sigma_l^z
\),
so that the disorder operators become nonlocal Pauli strings.

In our implementation, we define the lattice kink and antikink operators as
\begin{align}
{\cal K}(j) = \left( \prod_{k<j} \sigma_k^z \right) \sigma_j^+,\quad
{\cal A}(j) = \sigma_j^- \left( \prod_{k>j} \sigma_k^z \right).
\label{eq:K_A_ops}
\end{align}
The orientation of the string is a convention chosen such that composite operators ${\cal K}(j_1){\cal A}(j_2)$ produce a finite string between the kink ${\cal K}(j_1)$ at site $j_1$ and the antikink ${\cal A}(j_2)$ at site $j_2$, corresponding to a localized domain of flipped N\'{e}el order.

\subsection{Parameter dictionary and universality}

Equating the sine-Gordon couplings in Eqs.~\eqref{eq:SG_spin} and \eqref{eq:SG_QCD} yields a direct identification between the XXZ anisotropy and the number of colors~\cite{Korepin:1993kvr,Giamarchi:2003ooa} (see also Appendix~\ref{sec:continuum_mapping}),
$\beta^2 = \beta_{\rm QCD}^2$, hence $K(\Delta)=1/N$.

Matching the overall mass scales fixes
\begin{equation}
\sqrt{\lambda}\;\sim\;\frac{v(\Delta)}{a}\;\sim\;
2\pi J\,\frac{\sqrt{1-\Delta^2}}{\arccos\Delta},
\label{eq:Jlambda}
\end{equation}
up to a single nonuniversal constant.
Finally, the strength of the cosine deformation implies
\begin{equation}
\frac{m}{\sqrt{\lambda}}\;\propto\;\frac{h_s}{J}.
\label{eq:mhs}
\end{equation}

With these identifications, the soliton mass gaps \eqref{eq:gapXXZ} and \eqref{eq:gapQCD} take the identical universal form,
\begin{equation}
M_{\rm sol}
\;\propto\;
\Lambda
\left(
\frac{\delta}{\Lambda}
\right)^{\!\frac{1}{2-1/N}},
\end{equation}
where $(\Lambda,\delta)=(J,h_s)$ on the spin-chain side and $(\Lambda,\delta)=(\sqrt{\lambda},m)$ on the $\mathrm{QCD}_2$ side.
This matching provides a direct, quantitative correspondence between the staggered-field XXZ chain and massive large-$N$ $\mathrm{QCD}_2$ at the level of nonperturbative mass generation.
\smallskip

The soliton mass formulas derived from the sine-Gordon description are valid for arbitrary $N$, including the large-$N$ limit relevant for $\mathrm{QCD}_2$; the apparent restriction $N\le 2$ arises solely from realizing the sine-Gordon theory via a spin-$\tfrac12$ XXZ chain with a staggered longitudinal field, and reflects limitations of that specific lattice ultraviolet completion rather than of the continuum mapping itself.

\subsection{Soliton--antisoliton potential as an observable}
\label{sec:potential}

Our central observable is the effective interaction potential between a soliton and an antisoliton separated by a distance $R$. Operationally, we define this potential as the static energy cost of imposing a configuration with topological charges $+1$ and $-1$ centered at positions separated by $R$, relative to the vacuum. In the continuum SG description, this corresponds to minimizing the energy functional
\begin{equation}
E[\phi]
=
\int dx\,
\left[
\frac{1}{2}\,(\partial_x\phi)^2
+\frac{\alpha}{\beta^2}\Big(1-\cos(\beta\phi)\Big)
\right],
\label{eq:sg_energy_functional}
\end{equation}
subject to boundary conditions that enforce a soliton at $x=-R/2$ and an antisoliton at $x=+R/2$ (and overall $B=0$). We define the potential as 
\begin{equation}
V_{S\bar{S}}(R)
\equiv
E_{S\bar{S}}(R)
-2M_S,
\label{eq:potential_definition_continuum}
\end{equation}
where $M_s$ is the single-soliton rest energy (mass). By construction, $V_{S\bar{S}}(R)\to 0$ as $R\to\infty$, while at finite $R$ it captures the attractive interaction.

In summary, the problem of nucleon--antinucleon interactions in large-$N$ QCD$_2$ is mapped to soliton--antisoliton dynamics in the sine-Gordon model and, upon discretization, to kink--antikink physics in the XXZ chain. The primary observable is the separation-dependent interaction energy $V_{S\bar{S}}(R)$ defined by \eqref{eq:potential_definition_continuum}, which we will estimate on an IBM Quantum Nighthawk processor using qubit-native implementations of \eqref{eq:xxz_hamiltonian} and compute in the continuum limit using matrix product states.

\section{Quantum-Centric Implementation}
Our goal is to extract from quantum hardware the kink-antikink interaction potential, i.e, the lattice analog of the binding energy definition Eq.~(\ref{eq:potential_definition_continuum}).

Let $E_0$ denote the vacuum energy measured on the hardware, and define the single defect energy costs as $M_{\cal K}\equiv E_{\cal K}-E_0$ and $M_{\cal A}\equiv E_{\cal A}-E_0$. 
The lattice binding energy is then $V(r)=E_{\cal KA}(r)-E_0-M_{\cal K}-M_{\cal A}$, which expanded, is the measured energy estimator used for the hardware, 
\begin{equation} \label{eq:potential}
V(r) \equiv E_{\cal KA}(r) - E_{\cal K}(r) - E_{\cal A}(r) + E_0,
\end{equation}
where each $E_{\bullet}$ is an energy expectation value in a (generally non-eigen) state obtained from the vacuum by acting with the nonunitary disorder operators (\ref{eq:K_A_ops}) on the interacting quantum many-body ground state. 
Keeping $E_0$ explicit makes the estimator invariant under $H\to H+c$ and improves cancellation of hardware offsets.

Unlike conventional local and unitary operators, these operators contain extended string components and local ladder operators leading to nonunitary transition amplitudes. 
Such quantities can be challenging to sample efficiently with classical Monte Carlo due to the complex phase structure and normalization by operator dependent overlaps, but may be classically accessible with recently adapted tensor network methods. 
Additionally, utilizing quantum hardware provides a direct estimator. 

Complementary DMRG calculations at large $L$ (Sec.~\ref{sec:TN}) provide classical benchmarks and connect the lattice observable to a universal continuum scaling function.

\paragraph{Lattice model and parameter regime.}
For our quantum simulations we work with a $L=14$ site spin chain subject to open boundary conditions governed by Eq.~(\ref{eq:XXZ_staggered}). 
Throughout we use $(J_x,J_y,J_z)=(0.8,0.8,0.79)$ (i.e., $\Delta=J_z/J_x\simeq 0.9875<1$). 
A nonzero staggered field $h_s$ selects a N\'{e}el-like vacuum and localizes the solitons. 
We acquire hardware data at two field magnitudes, $|h_s|=0.10$ and $|h_s|=0.03$, running separate batches for the two signs $\pm h_s$ at each magnitude. 
In the main text we report the interaction potential extracted from the domain wall channel discussed above and average over the four translation related endpoint configurations (two central choices and two endpoint-movement conventions).
The complementary channel, as well as fully symmetrized averages that combine $\pm h_s$, are documented in Section~\ref{sec:supp_channels} as diagnostics of lattice/finite-size sensitivity and hardware systematics.

\paragraph{Ground state preparation.}
For each sign of $h_s$ we prepare an approximate interacting quantum ground state $|\psi_0(h_s)\rangle$ using a shallow variational quantum eigensolver (VQE) circuit. Our strategy is to design a circuit that contains enough expressivity needed to capture short-range and strongly-coupled physics but maintains a minimal two-qubit circuit depth. 
We find that two entangling layers (106 parameters for $L=14$) does this. The optimized parameters are then held fixed and used for all subsequent measurements at different separations $r$.

\paragraph{Nonunitary kink and antikink operators.}
Let ${\cal K}_j$ and ${\cal A}_j$ denote the kink and antikink operators, respectively, of Eq.~(\ref{eq:K_A_ops}). The relevant energy terms are then conditional expectation values of the form 
\begin{equation} 
E_O \equiv 
\frac{\bra{\psi_0} O^\dagger H O \ket{\psi_0}}
     {\bra{\psi_0} O^\dagger O \ket{\psi_0}},
\qquad O\in\{{\cal K,A,KA}\},
\end{equation}
along with $E_0 = \langle\psi_0|H|\psi_0\rangle$.
Because ${\cal K}_j$ and ${\cal A}_j$ are nonunitary, they are implemented using an ancilla-mediated circuit with projective measurement and postselection, described in Appendix~(\ref{sec:ancilla_string_implementation}). We then evaluate the four energies $\{E_0, E_{\cal K}, E_{\cal A}, E_{\cal KA}\}$ for each separation $r$ needed to construct V(r) in Eq.~(\ref{eq:potential}).

\paragraph{Configuration averaging and sublattice channels.}
Because we work on a finite open chain, the extracted interaction potential $V(r)$ has a weak dependence on where the kink and antikink insertion sites $(j_{\cal K}, j_{\cal A})$ are located due to boundary effects and lattice scale sensitivity.
To mitigate these geometry-dependent biases, we evaluate $V(r)$ in several embeddings around the middle of the chain. 
For $L=14$ we choose two neighboring central sites ($c=6$ and $c=7$) and two insertion site movement conventions (``${\cal K}$-first" and ``${\cal A}$-first"). This results in four different configurations that we average over. 
The spread between individual configurations provides an empirical estimate of the residual dependence of the embedding at fixed $L$. 

The staggered background field $h_s$ selects a staggered, N\'{e}el-like background, thereby distinguishing two inequivalent embeddings (channels) depending on how the kink and antikink insertion sites sit relative to the staggered order. 
Here in the main text we report on the interaction potential extracted from the “domain wall” channel, where the kink (antikink) predominantly flips a spin on the down (up) sublattice. 
This is consistent with the standard picture of a localized kink as a single domain wall between the two N\'{e}el vacua~\cite{Rutkevich_2018}. 

The complementary channel also yields a well defined postselected state at finite $h_s$, but constitutes a distinct lattice insertion channel that exhibits pronounced lattice-scale structure and does not show the same clean approach to the continuum soliton sector in our tensor network benchmarks\footnote{
In the classical N\'{e}el-state limit, such as for sufficiently strong staggered pinning, endpoint ladder operators acting on the ``wrong" sublattice annihilate the vacuum locally, $\sigma^+|\uparrow\rangle=0$ and $\sigma^-|\downarrow\rangle=0$, so this channel is absent. For the interacting ground state at finite $|h_s|$ studied here, quantum fluctuations make both endpoint embeddings nonzero, though they correspond to inequivalent channels at finite lattice spacing.}. We therefore treat it as a diagnostic and document it, together with the fully symmetrized averages over $\pm h_s$, in Appendix~\ref{sec:QC_WF_app}. 

Experimentally, all circuits are run on the quantum hardware within a single batch to minimize recalibration drift. 
Device details, error suppression settings, and ancilla readout calibrations are summarized in Appendix~\ref{sec:hardware_details}.

\section{Results}
\begin{figure*}[t]
\centering
\includegraphics[width=0.8\textwidth]{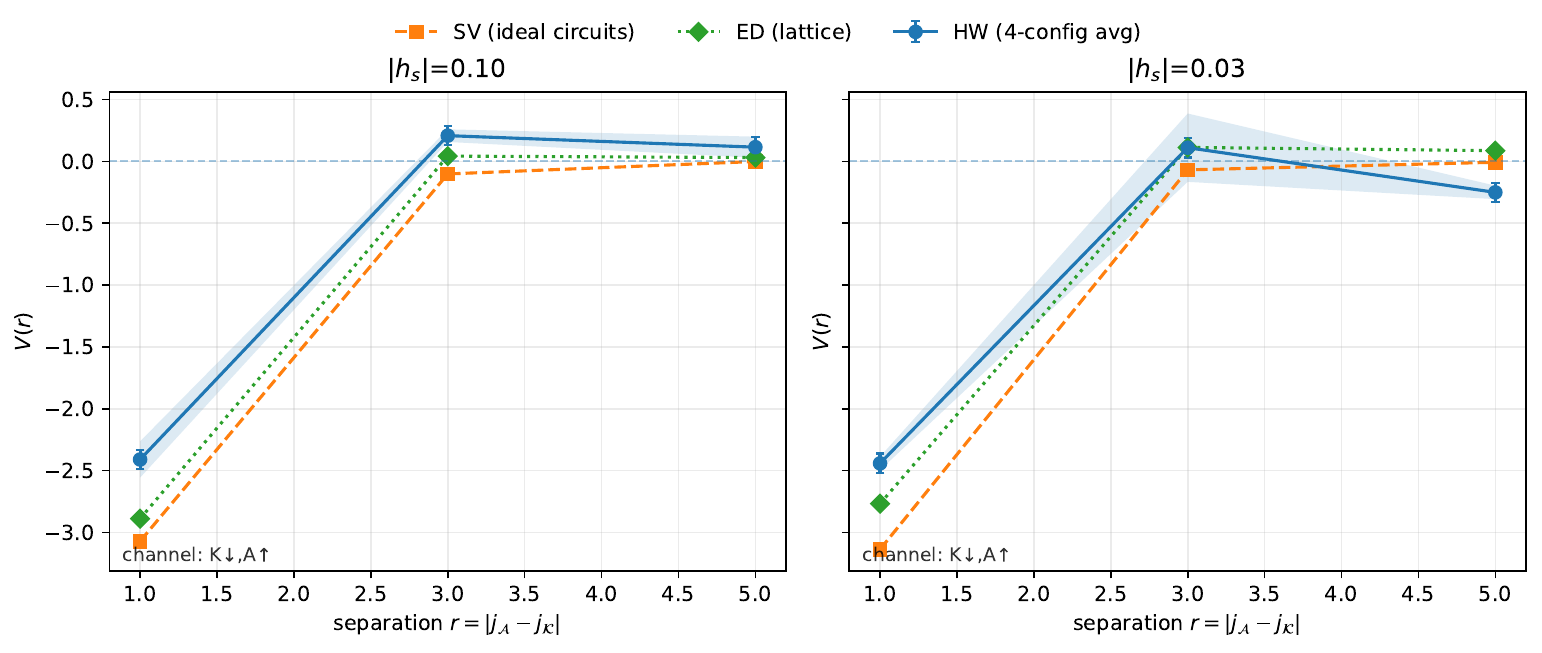}
\caption{Kink--antikink interaction potential $V(r)$ extracted on quantum hardware (HW) compared with exact diagonalization (ED) for an $L=14$ chain, shown for $|h_s|=0.10$ (left) and $|h_s|=0.03$ (right). The HW points show the 4-configuration average within the domain wall channel. Error bars denote the statistical uncertainty and the shaded band indicates the min/max spread across the contributing configurations. For $L=14$ this channel occurs at odd endpoint separations $r=|j_{\cal A}-j_{\cal K}|\in\{1,3,5\}$.} 
\label{fig:Vr}
\end{figure*}

Figure~\ref{fig:Vr} shows the interaction energy $V(r)$ between a kink and an antikink as a function of their separation $r$ for a $L=14$ site spin chain. 
For each $r$ we report the 4-configuration hardware averages and statistics (HW) alongside two classical references: (i) exact diagonalization of the lattice Hamiltonian (ED); (ii) the ideal statevector evaluation (SV) of the same circuits run on the hardware but with a classical and noiseless evaluation. 
Therefore the HW-SV difference isolates hardware noise whereas the SV-ED difference isolates residual error from using a shallow variational vacuum ansatz.

Across the two staggered field strengths, $|h_s| = 0.10$ and $|h_s| = 0.03$, the quantum hardware resolves a strongly attractive interaction at short range 
that becomes less negative at larger separations, consistent with the expected qualitative kink--antikink potential.
We note the qualitative potential shape is robust under the $\sim 3 \times$ change in $|h_s|$ and configuration averaging.
Quantitatively, the largest deviations between HW and ED occur at the smallest separations, where the problem is most sensitive to lattice discretization, finite-size effects, and to imperfections in the prepared vacuum and postselected excitation circuit.
The configuration spread (shaded band in Fig.~\ref{fig:Vr}) provides an empirical diagnostic of this residual geometry dependence on the finite chain and should be interpreted as a finite-size/lattice systematic and not purely statistical noise, which is accounted for in the error bars.

In Appendix~\ref{sec:QC_WF_app} we additionally present channel-resolved results that separate distinct sublattice embeddings of the local insertions, and  fully symmetrized averages that combine both signs $\pm h_s$ (an 8-configuration average), which serve as diagnostics of lattice artifacts and of residual hardware systematics.

\begin{figure}[t]
\centering
\includegraphics[width=0.85\columnwidth]{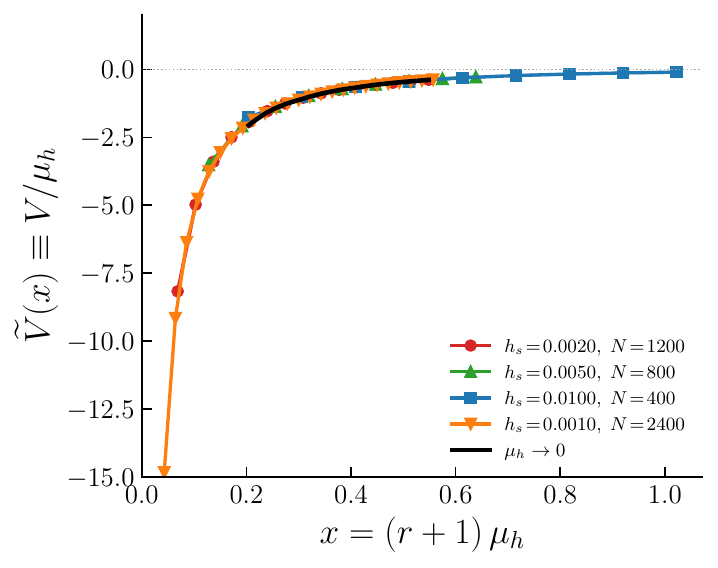}
\caption{Scaling collapse of the four primary large-$L$ DMRG datasets.
The $\mu_h\!\to\!0$ continuum extrapolation (solid black line) with
one-sigma uncertainty band (gray) is obtained by a linear-in-$\mu_h$
fit at each fixed $x$; the four field values agree to $1.7\%$ RMS
over the common overlap window.}
\label{fig:scaling_extrap}
\end{figure}

To connect the $L\!=\!14$ hardware results to the continuum field theory, we perform DMRG calculations at system sizes up to $L\!=\!2400$ and staggered fields down to $|h_s|=0.001$ (see Appendix~\ref{sec:TN} for full details).  
When the aligned-channel potential is rescaled as $\widetilde V = V(r)/\mu_h$ versus $x=(r\!+\!1)\mu_h$, where $\mu_h=(|h_s|/J)^{1/(2-K_{\rm BA})}$ is a dimensionless mass scale fixed by the Bethe-ansatz Luttinger parameter $K_{\rm BA}=K(\Delta)\approx 0.527$, the data from four field values ($|h_s|=0.01$, $0.005$, $0.002$, $0.001$; $L=400$--$2400$) collapse onto a single universal curve with $1.7\%$ RMS agreement (Fig.~\ref{fig:scaling_extrap}).  
A linear-in-$\mu_h$ extrapolation at each fixed scaled distance $x$ yields the continuum soliton--antisoliton scaling function $F_{\rm cont}(x)$, providing a controlled estimate of the continuum kink--antikink potential.

\section{Discussion}

Our results demonstrate that a kink--antikink interaction potential, defined through combinations of conditional energies involving nonunitary disorder operator insertions, can be resolved on current superconducting quantum hardware for a nontrivial interacting spin model.
In the main text we focus on the ``domain wall" channel, where the kink (antikink) insertions predominantly act on the down (up) sublattice relative to the staggered N\'{e}el-like order.
This channel most directly matches the standard lattice picture of a localized kink as a single domain wall interpolating between the two N\'{e}el-like vacua and exhibits the smoothest approach to large-$N$ behavior in complementary classical benchmarks.

The measured interaction potential displays the expected qualitative behavior of a strong short-range attraction which asymptotes to $V(r)\approx 0$ at larger separations (see Figs.~\ref{fig:scaling_extrap} and \ref{fig:Vr_supp8}).
This behavior is stable under substantial variation of the staggered field strength.
The shaded configuration-spread band in Fig.~\ref{fig:Vr} provides an empirical measure of a residual geometry dependence after mitigation and averaging. The spread is largest at the shortest separations where boundary effects, circuit depth, and postselection sensitivity are all maximal.

A useful diagnostic is the decomposition
\begin{align*}
\!\Delta_{\rm HW-ED}(r)
&=V_{\rm HW}(r)-V_{\rm ED}(r) \nonumber \\
&=\underbrace{\big[V_{\rm HW}(r)-V_{\rm SV}(r)\big]}_{\text{hardware bias}}
+\underbrace{\big[V_{\rm SV}(r)-V_{\rm ED}(r)\big]}_{\text{ansatz bias}},
\end{align*}
where $V_{\rm SV}(r)$ is the ideal statevector evaluation of the same circuits.
In our data the SV curves already deviate from ED at the smallest separations, indicating a residual bias from using a shallow variational vacuum ansatz rather than the exact ground state. The additional shift from SV to HW quantifies the impact of device noise and postselection on top of our ansatz circuit family. A full visualization of $V_{\rm HW}(r)$, $V_{\rm SV}(r)$, $V_{\rm ED}(r)$ and the corresponding bias terms is provided in Fig.~\ref{fig:Vr_decomp}.
Reducing $|h_s|$ decreases explicit symmetry breaking and can reduce certain lattice pinning effects, but it also increases finite-size sensitivity on a small fixed chain. Therefore, the dominant short-distance mismatch between HW and ED can not vanish simply by changing $|h_s|$.

Although absolute conditional energies on hardware can be strongly biased, the interaction potential is built as a difference of differences (Eq.~\ref{eq:potential}). This structure cancels a large fraction of state-independent offsets and slowly varying coherent errors, and is one reason a meaningful $V(r)$ can be extracted even when individual $E_{\bullet}$ energy estimates are imperfect.
The remaining systematics are dominated by accumulated two-qubit gate and decoherence errors in the ancilla-string circuits, ancilla readout errors on postselection, and  sensitivity to operator insertion points on a finite chain.
We mitigate these effects through batch execution to limit drift, dynamical decoupling, measurement twirling, explicit ancilla calibration, and configuration averaging, details of which are in Appendix~\ref{sec:QC_WF_app}.

A second important outcome of this work is the continuum extraction of the potential using matrix product states. The large-$L$ DMRG data show that the
potential has a controlled continuum limit in which it remains attractive. This shows that our conclusions drawn from the small system quantum simulation connect to continuum physics.

Several near term improvements could tighten agreement with ED at short distances: a deeper or more expressive vacuum ansatz, stronger error mitigation (e.g., noise amplification and extrapolation), and layouts that further reduce string depth and ancilla readout asymmetries.
A natural next step is to push the protocol used to longer chains to reduce finite-size artifacts and to more cleanly access the continuum scaling in the quantum simulation. The corresponding classical benchmark has been established by our tensor network simulation.
Nevertheless, the present protocol of combining variational state preparation with postselected nonunitary operator insertions provides a route to quantum computation of disorder operator observables and interaction energies, and can be generalized to larger lattices and to gauge theory inspired models as hardware improves.

\subsection*{Acknowledgments}
\noindent We thank Vladimir Korepin, Sergey Lukyanov and Martin Savage for insightful discussions. 
This work was supported in part
by the U.S. Department of Energy, Office of Science, Office of Nuclear Physics, Inqubator for Quantum Simulation (IQuS) under Award Number DOE
(NP) Award DE-SC0020970 (S.G.). 
This material is based upon work supported by the U.S. Department of Energy, Office of Science, National Quantum Information Science Research Centers, Co-design Center for Quantum Advantage (C2QA) under contract number DE-SC0012704.
This work was also supported by the U.S. Department of Energy, Office of Science, Office of Nuclear Physics, Grant No. DE-FG02-97ER-41014 (UW Nuclear Theory, S.G.) and DE-FG02-88ER40388 (SBU Nuclear Theory, D.K.). S.G. was supported in part by a Feodor Lynen Research fellowship of the Alexander von Humboldt foundation. This work was also supported, in part, by the Department of Physics and the College of Arts and Sciences at the University of Washington. CVC is supported in part by the RPI-IBM Future of Computing Research Collaboration.

\appendix

\section{Continuum mapping}
\label{sec:continuum_mapping}

\subsubsection{XXZ chain $\to$ sine-Gordon}

In the absence of the staggered field, the XXZ chain is described by a Luttinger liquid with Hamiltonian density
\begin{equation}
\mathcal H_0
=
\frac{v}{2}
\left[
K(\partial_x\theta)^2
+
\frac{1}{K}(\partial_x\phi)^2
\right],
\end{equation}
where the Luttinger parameter and velocity are known exactly from the Bethe ansatz~\cite{Luther:1975wr,Haldane:1981zza,Giamarchi:2003ooa,Korepin:1993kvr},
\begin{equation}
K(\Delta)=\frac{\pi}{2\bigl(\pi-\arccos\Delta\bigr)},
\qquad
\frac{v(\Delta)}{a}
=
2\pi J\,
\frac{\sqrt{1-\Delta^2}}{\arccos\Delta},
\label{eq:KvXXZ}
\end{equation}
with $a$ the lattice spacing.

The staggered field couples to the staggered magnetization,
\begin{equation}
\sigma_n^z
\;\sim\;
\cdots
+
c_1(\Delta)\,(-1)^n
\cos\!\bigl(\sqrt{4\pi K}\,\phi(x)\bigr)
+\cdots ,
\end{equation}
and generates a relevant sine-Gordon perturbation~\cite{Affleck1999,Oshikawa1997}. The resulting continuum Hamiltonian density reads
\begin{equation}
\mathcal H_{\rm SG}
=
\frac{v}{2}
\left[
K(\partial_x\theta)^2
+
\frac{1}{K}(\partial_x\phi)^2
\right]
-
g_s \cos(\beta\phi),
\label{eq:SG_spin}
\end{equation}
with
\begin{equation}
\beta^2 = 4\pi K(\Delta),
\qquad
g_s = C_s\,\frac{h_s}{a},
\end{equation}
where $C_s\sim O(1)$ absorbs the nonuniversal bosonization amplitude $c_1(\Delta)$ and normalization conventions.
Note that in this convention the cosine argument is $\beta\phi$ with $\beta^2=4\pi K$; some references instead define the SG coupling through $\tilde\beta^2/(8\pi)=1-\gamma/\pi$ with $\gamma=\arccos\Delta$, corresponding to a different field normalization. Throughout this work we use $\beta^2=4\pi K$ exclusively.
The scaling dimension of the cosine operator is $\beta^2/(4\pi)=K$, so the perturbation is relevant throughout the XXZ critical regime.

The sine-Gordon theory~\eqref{eq:SG_spin} dynamically generates a mass gap, corresponding to the soliton (kink) mass. Up to a nonuniversal prefactor, the gap scales as
\begin{equation}
M_{\rm sol}^{\rm (XXZ)}
\;\sim\;
\frac{v}{a}
\left(
\frac{g_s\,a}{v}
\right)^{\!\frac{1}{2-K}}
\;\sim\;
J\left(\frac{h_s}{J}\right)^{\!\frac{1}{2-K(\Delta)}} .
\label{eq:gapXXZ}
\end{equation}
The exponent $1/(2-K)$ is universal and entirely fixed by the anisotropy $\Delta$.

\subsubsection{Large-$N$ $\mathrm{QCD}_2$ $\to$ sine-Gordon}

The low-energy effective theory of large-$N$, single-flavor $\mathrm{QCD}_2$ in the ’t~Hooft limit $\lambda=g^2N$ can be written in bosonized form as a sine-Gordon theory,
\begin{equation}
\mathcal L_{\mathrm{QCD}_2}
=
\frac12(\partial_\mu\varphi)^2
+
\frac{\mu^2}{\beta_{\rm QCD}^2}
\cos(\beta_{\rm QCD}\varphi-\theta),
\label{eq:SG_QCD}
\end{equation}
with the exact identification
\begin{equation}
\beta_{\rm QCD}^2=\frac{4\pi}{N}.
\label{eq:betaQCD}
\end{equation}
The cosine amplitude is controlled by the quark mass $m$,
\begin{equation}
\mu^2 \sim C_\mu\, m\,\sqrt{\lambda},
\end{equation}
where $C_\mu$ depends on the normalization scheme but the scaling with $m$ and $\lambda$ is robust.
The vacua of the cosine lie at $\beta_{\rm QCD}\varphi=2\pi k+\theta$, so a single soliton (antisoliton) interpolates by $\Delta\varphi=\pm 2\pi/\beta_{\rm QCD}$. The topological baryon number is
\begin{equation}
B=\frac{\beta_{\rm QCD}}{2\pi}\bigl[\varphi(+\infty)-\varphi(-\infty)\bigr],
\label{eq:baryon_charge_QCD}
\end{equation}
which yields $B=\pm 1$ for a single soliton/antisoliton, identifying them with baryons and antibaryons.

The sine-Gordon theory~\eqref{eq:SG_QCD} also generates a soliton mass gap,
\begin{equation}
M_{\rm sol}^{(\mathrm{QCD}_2)}
\;\sim\;
\sqrt{\lambda}
\left(
\frac{m}{\sqrt{\lambda}}
\right)^{\!\frac{1}{2-\beta_{\rm QCD}^2/(4\pi)}}
=
\sqrt{\lambda}
\left(
\frac{m}{\sqrt{\lambda}}
\right)^{\!\frac{1}{2-1/N}} .
\label{eq:gapQCD}
\end{equation}

\section{Tensor network benchmarks and continuum scaling analysis}
\label{sec:TN}
\label{app:continuum-collapse}

The quantum hardware results of the main text demonstrate that the
kink--antikink potential can be resolved at $L\!=\!14$.  In this appendix we
benchmark those results against large-$L$ DMRG calculations~\cite{White:1992zz,Schollwoeck:2010uqf,Banuls:2018jag,Banuls:2019bmf}, characterize
finite-size artifacts in the $L\!=\!14$ data, and present the scaling analysis
that connects the lattice DMRG data to the continuum field theory.

\subsection*{Scaling variables}
\label{sec:scaling_variables}

At $h_s\!=\!0$, the XXZ chain with anisotropy $\Delta=J_z/J_x$ is described
by a Luttinger liquid.  The Bethe-ansatz Luttinger parameter is
\begin{equation}
K_{\rm BA}
=
\frac{\pi}{2\bigl(\pi-\arccos\Delta\bigr)} ,
\label{eq:K_BA}
\end{equation}
which gives $K_{\rm BA}=0.5265$ for $\Delta=0.79/0.8=0.9875$.

The staggered field $h_s$ perturbs this fixed point by coupling to the
staggered magnetization, an operator whose scaling dimension at the
Luttinger liquid fixed point is $K_{\rm BA}$.  Since $K_{\rm BA}<2$ the
perturbation is relevant, and the RG generates a mass gap
\begin{equation}
M \sim \left(\frac{|h_s|}{J}\right)^{\!1/(2-K_{\rm BA})} ,
\label{eq:mass_gap}
\end{equation}
where the exponent $1/(2-K_{\rm BA})$ is exact and follows directly from the
scaling dimension of the perturbation (see Eq.~\eqref{eq:gapXXZ}).  We define the dimensionless soliton mass
\begin{equation}
\mu_h
\equiv
\left(\frac{|h_s|}{J_x}\right)^{\!1/(2-K_{\rm BA})} .
\label{eq:mu_h}
\end{equation}
  The characteristic scaling length in lattice units
is $\xi\!=\!1/\mu_h$, proportional to the soliton Compton wavelength up to a nonuniversal prefactor; lattice simulations should be performed in a regime where $L/\xi \geq 20$ is large enough,
ensuring that the chains are well within the thermodynamic regime and the
kinks are spatially resolved on the lattice.

\subsection*{Finite-size effects and continuum extrapolation}
\label{sec:N14_diagnostics}

To assess the reliability of the $L\!=\!14$ quantum hardware results, we
overlay the $L\!=\!14$ DMRG data at $|h_s|=0.1$ and $|h_s|=0.03$ with the
large-$L$ continuum-regime data ($L=200$--$1200$, $|h_s|=0.002$--$0.01$) on
the same rescaled axes.

\begin{figure}[t]
\centering
\includegraphics[width=\columnwidth]{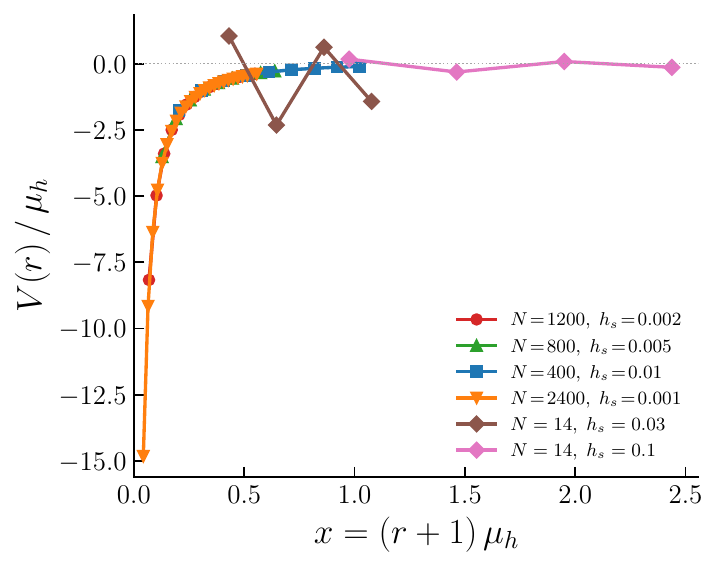}
\caption{Rescaled kink--antikink potential $V(r)/\mu_h$ versus
$x=(r+1)\mu_h$ for the four large-$L$ DMRG datasets (odd $r\!\geq\!3$, aligned channel) alongside
$L\!=\!14$ DMRG data at $|h_s|=0.03$ and $0.1$ (diamonds).
The large-$L$ data collapse onto a single universal curve; the $L\!=\!14$
curves show the finite-size structure expected for a 14-site chain,
consistent with the hardware data of Fig.~\ref{fig:Vr}.}
\label{fig:benchmark_collapse}
\end{figure}

Figure~\ref{fig:benchmark_collapse} shows the rescaled aligned-channel
potential $V(r)/\mu_h$ versus $x=(r+1)\mu_h$ [see Eq.~\eqref{eq:mu_h} for the definition of $\mu_h$] for odd separations
$r\!\geq\!3$.  The three large-$L$ curves collapse onto a single
universal curve (see Fig.~\ref{fig:benchmark_collapse}).  The $L\!=\!14$ data show
the attractive behavior seen on quantum hardware but exhibit finite-size
deviations from this curve.  At $|h_s|=0.03$, the $L\!=\!14$ curve oscillates
in sign across the accessible separations, reflecting the competition between
the kink--antikink binding energy and boundary corrections on a 14-site chain.
At $|h_s|=0.1$, the rescaled $L\!=\!14$ values are qualitatively smoother and
closer to the large-$L$ curve, consistent with the hardware data in
Fig.~\ref{fig:Vr}.

These comparisons establish two points relevant to the main text.  First,
the qualitative attractive shape of $V(r)$ seen on quantum hardware is
confirmed and placed in the context of the universal continuum curve by the
large-$L$ DMRG data.  Second, the quantitative values at $L\!=\!14$ are
subject to finite-size effects, particularly at small $|h_s|$ and short
separations.  Connecting to the continuum potential requires the large-$L$
DMRG benchmarks presented below, which control finite-size effects through
systematic volume checks.

We perform DMRG on the Hamiltonian of Eq.~\eqref{eq:XXZ_staggered} with
$(J_x,J_y,J_z)=(0.8,0.8,0.79)$.  The scaling analysis below uses data with four different values of the magnetization $h_s$. Three are each paired with a finite-volume check at a smaller system size; the fourth ($|h_s|=0.001$) has $L/\xi=25.7$, the largest ratio in the analysis, so finite-volume effects are expected to be negligible:
$ (L,\,|h_s|)=(400,\,0.01),$ $(200,\,0.01),$ $(800,\,0.005),$ $(400,\,0.005),$ $(1200,\,0.002),$ $ (800,\,0.002),$ $(2400,\,0.001).$
For the three smallest field values, the larger system in each pair is used as the large volume data and the smaller member serves as a finite-volume check.  The DMRG calculations use 15 sweeps with bond dimensions up to 400, and the ground-state energy is
converged to better than $10^{-10}$ for all system sizes shown.

The continuum soliton--antisoliton potential at scaled distance $x$ is obtained
in the limit $h_s\!\to\!0$, $r\!\to\!\infty$, $L/\xi\!\to\!\infty$ at fixed
$x\!=\!(r+1)\,\mu_h$.  On a finite lattice we define the rescaled
variables
\begin{equation}
x = (r+1)\,\mu_h ,
\qquad
\widetilde V(x) = \frac{V(r)}{\mu_h} ,
\label{eq:collapse_variables}
\end{equation}
where $r$ is the integer site separation between the kink and antikink
insertion points.  The effective distance $r+1$ accounts for the fact that a
lattice domain wall occupies one bond (two adjacent sites), so the geometric
center-to-center distance between a kink--antikink pair at separation $r$
sites is $r+1$ lattice spacings. Replacing $r+1$ by $r$ or $r+\tfrac12$ changes only the short-distance corrections and does not alter the continuum trend.

We restrict to the aligned channel (kink on the down sublattice, antikink on
the up sublattice) and odd separations $r\!\geq\!3$.  (excluding the contact point
$r\!=\!1$).  The
rescaling in Eq.~\eqref{eq:collapse_variables} uses the theory-fixed
Bethe-ansatz exponent determined by $K_{\rm BA}$ and a fixed convention for the nonuniversal mass scale.

Figure~\ref{fig:scaling_extrap} shows the rescaled potential
$\widetilde V(x)$ versus $x$ for the four field values.  The data
collapse onto a single curve.  The relative root-mean-square spread between the
four curves in the window $x\!\in\![0.20,\,0.55]$ is $1.7\%$ for $r_{\rm min}\!=\!3$ (and $1.4\%$
for $r_{\rm min}\!=\!5$).

Within each $(L,|h_s|)$ pair, the raw
aligned-channel potentials $V(r)$ agree at the sub-percent to percent level:
\begin{center}
\begin{tabular}{cccr}
\hline
$|h_s|$ & $L_{\rm partner}$/$L_{\rm main}$ & $L/\xi$ range & rel.\ RMS \\
\hline
$0.002$ & $800$/$1200$ & $13.7$--$20.6$ & $0.05\%$ \\
$0.005$ & $400$/$800$  & $12.8$--$25.5$ & $0.13\%$ \\
$0.01$  & $200$/$400$  & $10.2$--$20.4$   & $1.07\%$ \\
\hline
\end{tabular}
\end{center}
The finite-volume error decreases rapidly with $L/\xi$.  For the two smaller
field values the partner agreement is well below the inter-field spread in
the collapse, confirming that the residual drift in Fig.~\ref{fig:scaling_extrap}
is not due to finite-volume effects.

The scaling collapse of Fig.~\ref{fig:benchmark_collapse} demonstrates that the thermodynamic limit is reached, but a rigorous continuum potential requires the
additional step of extrapolating $\widetilde V(x,\mu_h)$ to $\mu_h\!\to\!0$
at each fixed $x$.

At each fixed scaled distance $x$, the rescaled potential receives lattice
corrections. We assume their form as
\begin{equation}
\widetilde V(x,\mu_h)
= \frac{V(r,\mu_h)}{\mu_h}
= F_{\rm cont}(x)
+ c_1(x)\,\mu_h + O(\mu_h^2) ,
\label{eq:lattice_correction}
\end{equation}
where $F_{\rm cont}(x)$ is the continuum scaling function and the leading lattice
correction is linear in $\mu_h$.  With four field
values, a linear fit in $\mu_h$ at each $x$ has two degrees of freedom,
providing a nontrivial test of the linearity assumption.

\begin{figure}[t]
\centering
\includegraphics[width=\columnwidth]{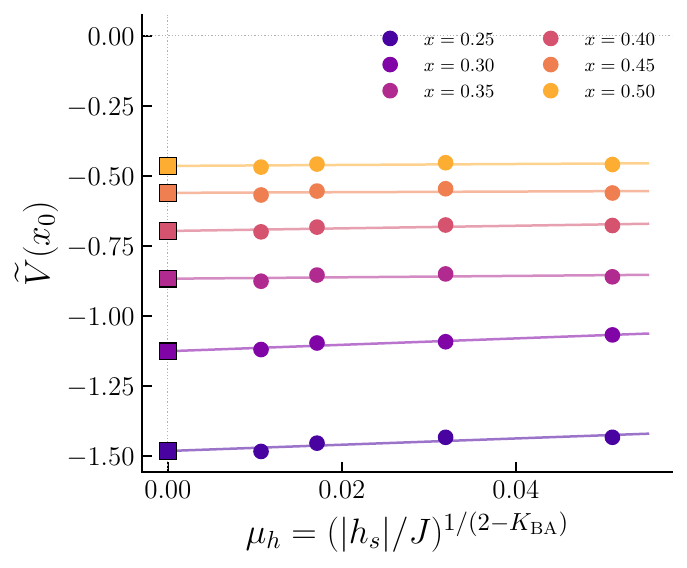}
\caption{Rescaled potential $\widetilde V(x_0)$ at fixed scaled
distances $x_0$, plotted against $\mu_h$ for the four field
values.  Linear fits (lines) extrapolate to $\mu_h\!=\!0$ (filled
squares), giving the continuum scaling function $F_{\rm cont}(x_0)$.
The monotonic dependence is consistent with a leading-order lattice
correction $\propto\mu_h$; the residuals of the linear fit provide
an estimate of the remaining lattice corrections.}
\label{fig:fixed_x_drift}
\end{figure}

Figure~\ref{fig:fixed_x_drift} shows $\widetilde V(x_0)$ at several fixed scaled
distances $x_0$, plotted against $\mu_h$ for the four field values.
Linear fits extrapolate to $\mu_h\!=\!0$ (filled squares), giving the continuum
scaling function $F_{\rm cont}(x_0)$.  The data is monotonic and well described
by a linear correction consistent with Eq.~\eqref{eq:lattice_correction}; the
residuals are small compared to the signal, supporting the assumed functional form.

\begin{figure*}[t]
\centering
\includegraphics[width=\textwidth]{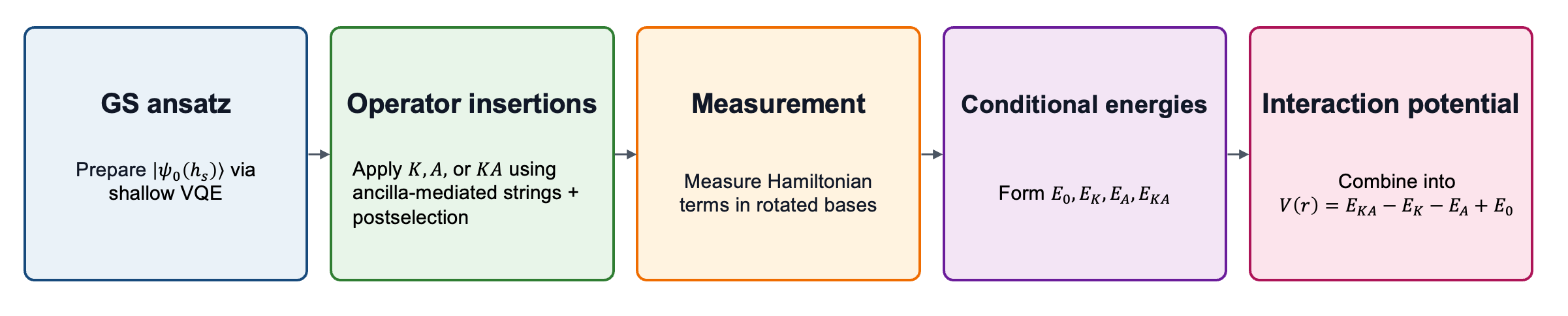}
\caption{A summary of the quantum-centric workflow used in this work.}
\label{fig:workflow}
\end{figure*}

\section{Quantum-Centric Workflow}
\label{sec:QC_WF_app}

Here we detail the steps in our quantum-centric workflow. 
A summary can be found in Figure~\ref{fig:workflow}

\subsection{Ground State Ansatz Construction}
\label{sec:GS_prep}

For each choice of staggered field sign $\pm h_s$, we prepare an approximate vacuum $|\psi_0(h_s)\rangle$ using a shallow parameterized circuit and a classical optimizer.  
The ansatz is chosen to balance two competing requirements to give the best final result possible. 
Namely, the ansatz needs to being expressive enough to capture the dominant staggered order at the lattice sizes accessible to the quantum hardware but remain sufficiently shallow to allow the subsequent postselected disorder operator circuits to execute with minimal noise. 

In practice we use a Hamiltonian inspired ansatz with two entangling layers acting on an initial N\'eel product state. We then optimize its parameters with a noise free classical simulator.  
The resulting parameter vectors are saved and reused to generate the ideal statevector reference curves and used to prepare the input state for all hardware circuits at the corresponding $\pm h_s$.

\subsection{Postselected disorder operator insertions}
\label{sec:ancilla_string_implementation}

The kink (${\cal K}$) and antikink (${\cal A}$) disorder operators used in the main text are nonunitary (see Eq.~(\ref{eq:K_A_ops})).  
To implement these nonunitary operators, we embed them into a larger unitary circuit with ancilla qubits, followed by projective measurement of the ancillas and postselection on the all zeros outcome.
Concretely, each state needed to calculate the kink-antikink interaction potential ($|\psi_0\rangle$, ${\cal K}|\psi_0\rangle$, ${\cal A}|\psi_0\rangle$, and ${\cal KA}|\psi_0\rangle$) is created by a dedicated circuit that first prepares the variational vacuum, then applies an ancilla-mediated string to implement the desired nonunitary operator insertion, and finally, measures the local Hamiltonian terms needed to estimate the corresponding conditional energy.

To estimate each energy we measure all single- and two-site Pauli observables appearing in the XXZ Hamiltonian by repeating the same circuit under three measurement bases ($X$, $Y$, and $Z$ rotations on the data qubits).
For each configuration and separation $r$, this yields a fixed set of circuits per energy component. 
The interaction potential is then calculated from the difference of differences in Eq.~(\ref{eq:potential}).  
Crucially, this difference of differences structure suppresses many state-independent offsets and slowly varying coherent errors that would otherwise strongly bias absolute energy estimates.

\subsection{Configuration averaging and sublattice channels}
\label{sec:supp_channels}

\paragraph{Four-configuration averaging.}
On a finite open chain, observables depend on where the kink and antikink insertions $(j_K,j_A)$ sit relative to the boundaries, as well as the hardware embedding.  
To reduce this geometry dependence, we evaluate $V(r)$ in multiple configurations that differ only by the choice of insertions along the chain, and then average.
Specifically, for $L=14$ we use two nearby center choices ($c=6$ and $c=7$) and two insertion update conventions (``${\cal K}$-first'' and ``${\cal A}$-first''), yielding four configurations: \texttt{offset0\_kfirst}, \texttt{offset1\_kfirst}, \texttt{offset0\_afirst}, and \texttt{offset1\_afirst}.
For a given $r$, we average over the subset of configurations that realize that separation without pushing either insertion point outside of the chain. 

\paragraph{Domain wall channel selection.}
On a staggered background, translation by one lattice site exchanges the two N\'eel-like sublattices.  
Consequently, local kink/antikink operator insertions  organize into distinct sublattice channels depending on whether $K$ acts predominantly on a down sublattice site and $A$ on an up sublattice site (or vice versa), relative to the staggered order of the vacuum.
In the standard picture of a localized kink as a single domain wall between two N\'eel-like vacua, a kink state contains exactly one ferromagnetic bond (two adjacent spins aligned) separating the two antiferromagnetic domains~\cite{Rutkevich_2018}.
In our finite-size setting, we identify the corresponding ``domain wall" channel by the signs of the local magnetizations at the insertion points in the exact ground state.
That is, we classify each configuration $(j_K,j_A)$ by $z_K=\langle Z_{j_K}\rangle$ and $z_A=\langle Z_{j_A}\rangle$ and select the channel with $z_K<0$ and $z_A>0$ (denoted $K\downarrow,A\uparrow$).
For $L=14$ this channel occurs at odd insertion point separations. Combined with simulating separations $r=|j_A-j_K|=0,\dots,6$ on the quantum hardware, the main text plot thus reports only the odd separations $r\in\{1,3,5\}$.

The complementary channel ($K\uparrow,A\downarrow$) and configurations where both insertion points lie on the same sublattice remain well-defined postselected states for $h_s\neq 0$ (because the vacuum has quantum fluctuations and is not a perfect N\'eel product state).  However, at the small lattice sizes accessible to present day quantum hardware utilized in this work, this additional channel exhibits enhanced sensitivity to lattice parity and boundary effects, which  introduces additional short distance structure and oscillations. 
Note that in the tensor network benchmarks shown in Section~\ref{sec:TN} this structure converges more slowly with increasing $N$.
For transparency and as a hardware diagnostic, we document fully symmetrized averages that include all configurations in the next paragraph, and reserve the domain wall channel for the main text, which is more appropriate as a discrete estimator of the continuum.

\begin{figure*}[t]
\centering
\includegraphics[width=\textwidth]{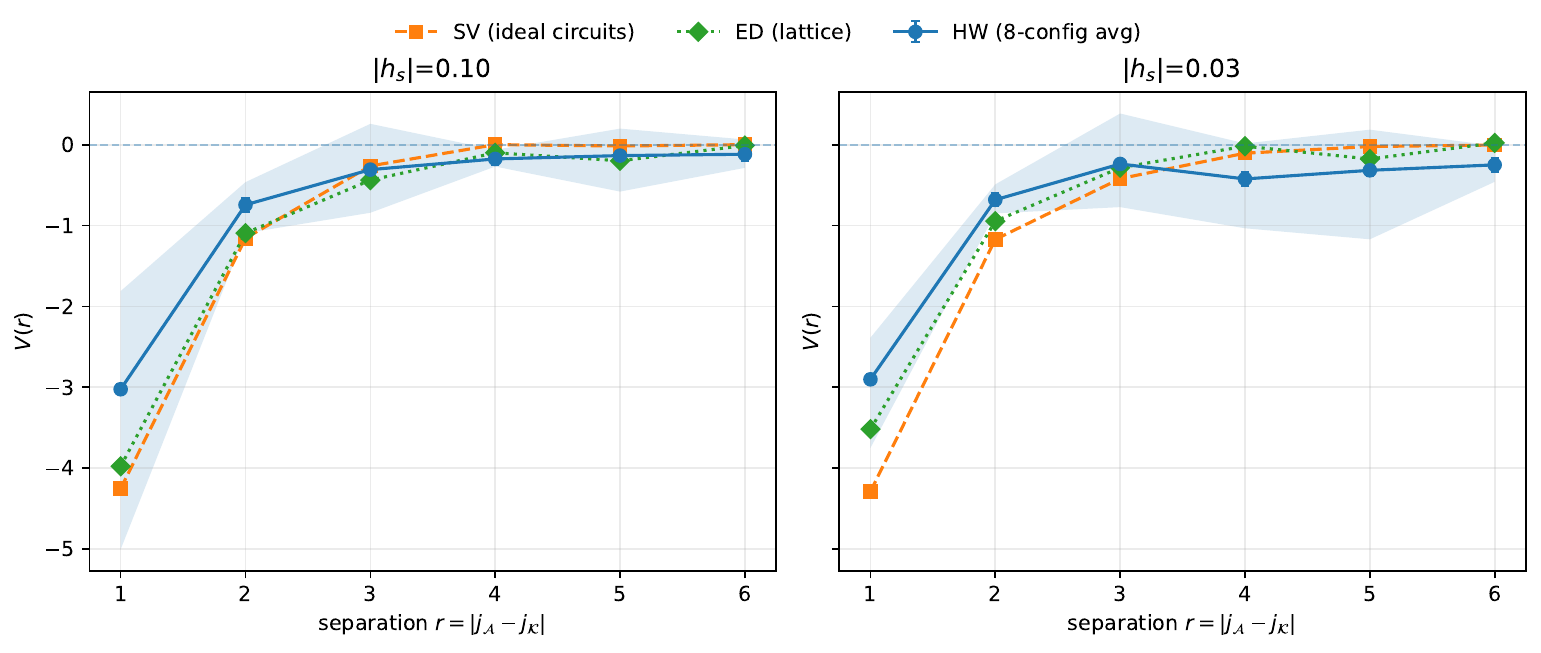}
\caption{Fully symmetrized kink--antikink interaction potential $V(r)$ on hardware (HW) compared with exact diagonalization (ED), for $|h_s|=0.10$ and $|h_s|=0.03$ on an $L=14$ chain.  The HW points show the 8-configuration average over two centers ($c=6,7$), two endpoint-update conventions (K-first/A-first), and both signs $\pm h_s$.
The shaded band indicates the min/max spread across the contributing configurations at each $r$.
This figure is provided as a diagnostic of residual lattice/geometry sensitivity and of the effect of fully symmetrized averaging; the main text focuses on the domain wall channel (Appendix~\ref{sec:supp_channels}).}
\label{fig:Vr_supp8}
\end{figure*}

\paragraph{Eight-configuration (fully symmetrized) averages.}
Given the above discussion, the ``8-configuration'' average is obtained by combining the four configurations above with both signs of the staggered field, $\pm h_s$. 
This averaging is equivalent to swapping the roles of the kink and the antikink under a one site translation\footnote{Equivalently, under the discrete transformation that exchanges the two N\'eel-like sublattices.} and can partially restore the corresponding symmetry at finite $L$.
It also reduces the purely statistical error bars by approximately $\sqrt{2}$ since the $\pm h_s$ datasets have comparable shot counts.
Figure~\ref{fig:Vr_supp8} shows the resulting symmetrized hardware averages alongside the exact diagonalization reference, together with the min/max spread across individual configurations which serve as an empirical measure of residual geometry dependence.

\begin{figure*}[t]
\centering
\includegraphics[width=\textwidth]{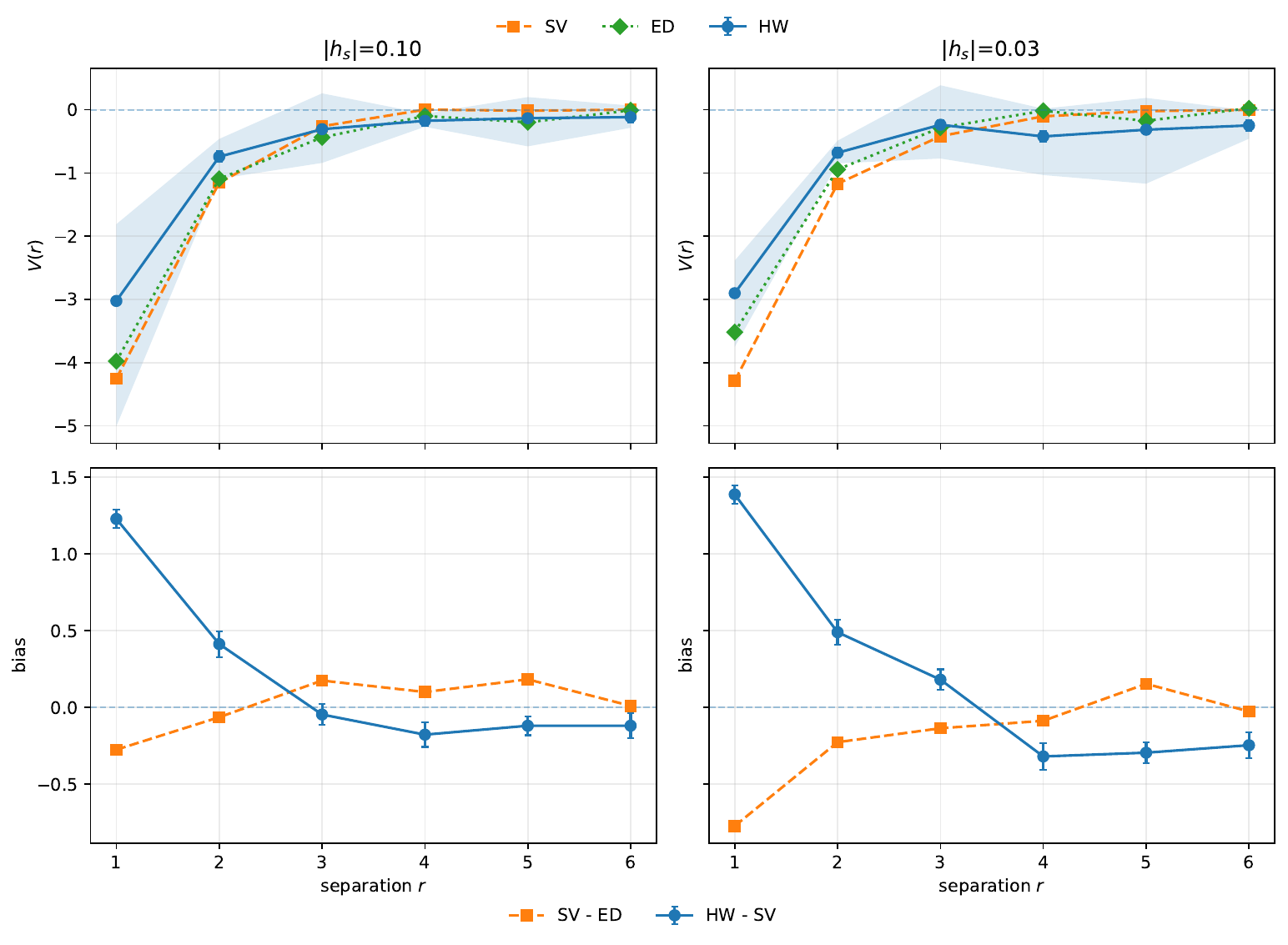}
\caption{Visualization of the diagnostic decomposition
$\Delta_{\rm HW-ED}(r)=\big[V_{\rm HW}(r)-V_{\rm SV}(r)\big]+\big[V_{\rm SV}(r)-V_{\rm ED}(r)\big]$.
Here $V_{\rm SV}(r)$ is the ideal statevector evaluation of the same circuits executed on hardware, and $V_{\rm ED}(r)$ is the exact diagonalization benchmark.
Top row: $V(r)$ for HW, SV, and ED using the fully symmetrized 8-configuration average.
Bottom row: the two bias contributions, $V_{\rm HW}-V_{\rm SV}$ (hardware noise and finite-shot sampling) and $V_{\rm SV}-V_{\rm ED}$ (ground state variational ansatz).}
\label{fig:Vr_decomp}
\end{figure*}

\subsection{Error suppression and ancilla readout calibration}
\label{sec:ancilla_calib}

All hardware circuits were executed within a single batch per dataset to reduce the impact of calibration drift.
Compilation used a hardware-aware qubit layout focused on qubit locality and quality metrics. Dynamical decoupling was enabled on idle windows to suppress dephasing during the nonlocal ancilla string operations.
We also enabled measurement twirling on the data qubits to reduce basis dependent readout biases.

Because ancilla outcomes determine postselection, ancilla readout errors are especially important.  Therefore, each batch includes a dedicated ancilla calibration job that prepares $|0\rangle$ on each ancilla and measures it repeatedly, yielding empirical confusion probabilities $P(0|0)$ and $P(1|0)$.
In post processing we apply a simple linear correction to the postselected probabilities using these confusion probabilities. 
For the data reported here, this correction produces a tiny quantitative shift and does not change the qualitative trends.

\subsection{Postselection and effective sample sizes}
\label{sec:postselection}

Compared to the raw shot budget, ancilla postselection reduces the effective sample size contributing to the conditional energies in Eq.~(\ref{eq:potential}).
Table~\ref{tab:resource_postselection} summarizes the hardware resources and postselection yields for both datasets.
We report the average postselection keep fractions for the single insertion circuits ($K$ and $A$), defined as $f_{\rm single}=\tfrac{1}{2}(f_K+f_A)$, and the double insertion circuits ($KA$), defined as $f_{\rm KA}$.
The quoted ranges indicate the minimum and maximum keep fractions observed across configurations, separations, and measurement bases.
In our datasets, $f_{\rm KA}$ is lower and exhibits a wider spread across circuits (configurations, separations, and measurement bases), consistent with deeper circuits and a larger number of postselected ancillas for the joint insertion circuits.
All statistical error bars shown in the figures are computed from the retained postselected samples and propagated through Eq.~(\ref{eq:potential}) under the assumption of independent shot noise between circuit groups.

\begin{table*}[t]
\caption{Hardware resources and postselection yields for the two datasets.
``Prod.~circs.'' counts the production (physics) circuits submitted to hardware, excluding ancilla calibration circuits.
Each dataset uses $10^4$ shots per production circuit, yielding a total raw shot budget of $3.66\times 10^6$ shots.
Because the ``A-first'' convention is only sampled at odd separations in our implementation, the total number of production circuits is 366.
The effective postselected shots are computed as the sum over circuits of $(\text{shots per circuit})\times(\text{keep fraction})$, separately for the single-insertion and joint-insertion circuit families.}
\label{tab:resource_postselection}
\begin{ruledtabular}
\begin{tabular}{ccccccc}
$|h_s|$ & shots/circ. & prod.~circs. & circs./$(r,\mathrm{cfg})$ & $\langle f_{\rm single}\rangle$ (min--max) & $\langle f_{\rm KA}\rangle$ (min--max) & eff.~shots (single / $KA$) \\
\hline
0.10 & $10^4$ & 366 & 9 & 0.502 (0.191--0.789) & 0.319 (0.047--0.790) & $1.20\times10^6$ / $3.83\times10^5$ \\
0.03 & $10^4$ & 366 & 9 & 0.483 (0.219--0.750) & 0.285 (0.053--0.713) & $1.16\times10^6$ / $3.41\times10^5$ \\
\end{tabular}
\end{ruledtabular}
\end{table*}

\section{Quantum Hardware Details}
\label{sec:hardware_details}

\subsection{Device}

All production results reported were obtained on the IBM Quantum Computer \texttt{ibm\_miami} featuring the Nighthawk superconducting processor and using the Qiskit Runtime batch execution mode.  
Within each batch, all circuits share a common compilation stack and were executed in a short time window to mitigate drift.

\subsection{Embedding: data-chain and ancillas}

We embed the $L=14$ logical spin chain onto a contiguous path of physical qubits (the data chain) and place a small set of ancilla qubits adjacent to this path to implement the nonunitary disorder operator strings.
We used a simple heuristic involving qubit locality and timing errors for score and choose different physical qubit chains. 
The the physical qubit embedding we chose was:

\begin{itemize}
\item Data chain: \texttt{[26, 25, 24, 34, 44, 54, 53, 43, 33, 23, 22, 21, 11, 12]}.
\item Ancillas: \texttt{[13, 32, 35, 42, 45, 52, 55]}.
\end{itemize}

Each ancilla is chosen to be directly connected to (at least) one nearby data qubit so that the controlled operations used in the string construction can be implemented with minimal SWAP overhead.

\subsection{Batch ancilla calibration}

For the same batch, the ancilla calibration circuits yield an average $P(0|0)\approx 0.98$ and $P(1|0)\approx 0.02$ across the seven ancillas used for postselection.  
These values are used in the ancilla readout correction described in Appendix~\ref{sec:ancilla_calib}.

\bibliography{main}

\end{document}